\documentclass{article}

\usepackage{graphicx}
\usepackage[verbose=true,letterpaper]{geometry}
\AtBeginDocument{
  \newgeometry{
    textheight=9in,
    textwidth=6.5in,
    top=1in,
    headheight=14pt,
    headsep=25pt,
    footskip=30pt
  }
}

\def\keywordname{{\bfseries \emph Keywords}}%
\def\keywords#1{\par\addvspace\medskipamount{\rightskip=0pt plus1cm
\def\and{\ifhmode\unskip\nobreak\fi\ $\cdot$
}\noindent\keywordname\enspace\ignorespaces#1\par}}

\renewenvironment{abstract}
{
  \centerline
  {\large \bfseries \scshape Abstract}
  \begin{quote}
}
{
  \end{quote}
}

\usepackage[utf8]{inputenc} 
\usepackage[T1]{fontenc}    
\usepackage{hyperref}       
\usepackage{url}            
\usepackage{booktabs}       
\usepackage{amsfonts}       
\usepackage{nicefrac}       
\usepackage{microtype}      

\title{Computational modeling to determine the physical characteristics of biological tissues for medical diagnosis \footnote{Published: Polyakov M. V. Computational modeling to determine the physical characteristics of biological tissues for medical diagnosis // International Journal of Engineering Systems Modelling and Simulation. — 2020. — Vol. 11, no. 4. — pp. 214-221. DOI: \url{https://doi.org/10.1504/IJESMS.2020.111276}}}

\author{
  Maxim V.~Polyakov \\
  Department of Information Systems and Computer Modeling\\
  Volgograd State University\\
 Volgograd, Russia \\
  \texttt{m.v.polyakov@volsu.ru} \\
}
\date{}

\begin{document}
\maketitle{}

\begin{abstract}
Timely diagnosis of breast cancer is an important task. This type of breast cancer is one of the most common diseases. The method of microwave radiothermometry is a promising direction for solving this problem. The method is based on measuring internal temperature of biological tissues in microwave frequency range. Computer simulations are used to improve the quality of diagnostics. Computer models make it possible to evaluate the effect of heat release in a malignant tumor on the thermal dynamics inside the mammary gland. It is necessary to build personalized models, taking into account the individual nature of the internal structure of the mammary gland in each patient. One of the problems is the determination of biophysical characteristics of biological components. Methods for determining these characteristics using computer simulations are proposed. The coefficient of thermal conductivity and specific heat of biological tissues are determined from known temperature distributions. Finding the physical parameters for a quasihomogeneous biological tissue is the first approximation for solving this problem. The least squares method is used as a solution method. The results obtained are in good agreement with previously known exact solutions, which indicates the applicability of this method for solving this class of problems. The efficiency of using parallel technologies in solving the inverse problem is investigated and the applicability of Open MP technology is demonstrated.
\end{abstract}


\section{Introduction}
The current level of development of computer technology allows us to significantly expand the class of applied problems which can be solved using simulation methods. Many scientific and technical problems which have been solved analytically are nowadays solved by numerical methods using relevant software for engineering analysis. In experimental studies of transient thermal processes, it is sometimes impossible to conduct direct measurements of required physical quantities, and these characteristics are inferred from the results of indirect measurements. The only way to find the required physical quantities for such problems is to solve the inverse boundary problems for heat conduction.

There are a number of applied studies in which it is impossible to determine the initial conditions. The mathematical models of such problems have the form of inverse boundary value problems with unknowns
initial data.

An important task is to determine the biophysical parameters of biological tissues. For example, in \cite{sarvazyan} the laws of the propagation of various types of elastic waves in biological tissues in the range of acoustic frequencies are investigated theoretically and experimentally. The contributions of imaginary and real components of the complex modulus elasticity to the speed of elastic waves is analyzed. It is shown that in soft tissues, low-frequency elastic disturbances propagate mainly as transverse waves.

The method of radio microwave thermometry is based on the measurement of tissue radiation in the ultrahigh range. An important role is played by methods for assessing the characteristics of emissivity biological tissues at microwave frequencies \cite{gupta}. Detailed mechanisms of heat transfer in tissues of living creatures and thermal response to cauterization were analized by \cite {liu} using the theory of biothermal transfer. The study was conducted experimentally using an infrared camera and thermocouples.

The article \cite {khokhlova} conducted a numerical simulation of thermal processes in biological tissues under the influence of focused ultrasound waves. The temperature distribution was calculated using inhomogeneous heat equation with a relaxation term, and an acoustic field was specified as a focused Gaussian beam. It was shown that an increase in intensity and a decrease in the exposure time at a constant dose of radiation, characteristic for therapeutic conditions, contributes to a significant localization and efficiency of the heating process.

The article \cite{kosin} showed a possibility of creating devices on the ArduinoUno platform for measuring temperature and electrical impedance of affected areas of biological tissues. The creation of a device for static and dynamic research in soft biological tissues (in particular, AAA tissue) is described by \cite{antonova}. In this paper, the preservation of viability of drug during the experiment, and for obtaining experimental data used to model aneurysmal stability and deformation, is considered. The authors rated the technical data of the device and evaluated and obtained repeatable results. The device allowed the authors to test not only arterial tissue, but also any soft biological tissues or artificial materials for medical applications in static and dynamic conditions.

One of urgent problems in medicine is to improve the quality of early diagnosis of breast cancer. Tumors should have a diameter of not more than 5-7 mm in order to detect them in a timely manner. However, according to statistics, the average size of newly detected tumors is significantly larger (1.34 cm), and the frequency of detection of tumors
up to 1 cm in diameter is 10-20 \%. Traditionally used methods do not allow us to detect a tumor at an early stage.
The global trend of modern mammology is the search for acceptable non-invasive methods for conducting preventive examinations.
To determine the effect of cancer on the background temperature, computational experiments based on the solution of the heat equation for biological tissues are traditionally carried out. Numerical assessment of thermal behavior of two types breast cancer:
ductal carcinoma and invasive ductal carcinoma, was performed by \cite{gomes}. Their analysis was based on taking into account two-dimensional geometry
which characterizes the anatomy of the breast. Thermal behavior was obtained using a numerical solution to the biothermal equation.
The design and experimental characteristics of a miniature radiothermometry system for measuring average volumetric temperature of tissue sites located at a depth of 5 cm in a body are presented by \cite{stauffer}. Their results show that this device allows us to accurately assess deep temperatures, and monitor for a long time.

An important indicator of the state of human tissues is temperature, especially at a considerable depth from the surface. It is possible to obtain additional and extremely important information about the condition of internal tissues and skin integuments by measuring the intrinsic radiation of biological tissues in microwave and infrared wavelength ranges. This is the basis for the diagnosis of diseases using the method of microwave radiothermometry. The method of microwave thermometry was first described by \cite{barrett}. Later it was proven that heat release of a tumor is directly proportional to its growth rate \cite{gautherie1}, \cite{gautherie2}. Therefore, microwave radiothermometry has unique ability to detect primarily rapidly growing tumors.

A new flexible hexagonal patch antenna with a substrate operating in the industrial, scientific and medical bands was presented by \cite{sheeba}. A two-step methodology was recommended. This implies the construction of an antenna with a hexagonal patch with a hexagonal slot and without a slot, into which the substrate was inserted, then the antenna was placed on skin of chest of human body model. Simulated antenna results show the location of a tumor present in skin and mammary gland this was achieved  by observing changes in density of skin and mammary gland with and without a tumor.

The RTM method is known for its simplicity and the possibility of wide application without medical restrictions, in contrast to various complex and refined approaches, such as highly selective screening of circulating tumor cells using microfluidic devices \cite{shirmohammadli}, magnetic-acoustic electric tomography \cite{zaheditochai}, genomic testing \cite{rotter}, development of fluorescent oligonucleotide probes for specific detection of a cancer marker \cite {saady}. The problems of choosing the most appropriate methods for diagnostics of breast cancer are actively discussed in the scientific literature e.g. \cite{yahyazadeh}. This discussion takes into account known shortcomings of traditional mammography. 3D imaging using various methods of three-dimensional imaging of the mammary gland were discussed by \cite{sundell}. Machine learning methods are increasingly used to diagnose medical images, including those for breast cancer \cite{bressan}.
Active application of the method of microwave radiothermometry in Russia began with the development of the microwave radio thermometer RTM-01-RES in 1997. The article \cite{vesnin} presented a systematic analysis of the data the on the role of microwave thermometry in risk assessment, diagnosis of breast pathology and in assessing effect of neoadjuvant therapy in treatment of breast cancer available in the literature. Various aspects of the application of microwave thermometry in breast diseases were described. These include diagnostic potential of the method and its importance in the differentiation of hyperplasia, benign and malignant diseases. Studies have also shown the prognostic role of microwave thermometry and its possible application for assessing the effect of preoperative chemotherapy for locally advanced breast cancer.

Important advantages of the microwave radiothermometry method are its ability to timely detect diseases and its absolute harmlessness, both for the patient and for medical personnel \cite{sedankin}. The use of microwave thermometry for diagnosis of breast cancer was investigated by \cite{losev}.

It is worth noting that breast cancer is not a specific problem for women. Statistics for this disease in women as a whole does not improve \cite{sharma}, but this disease is also observed in men \cite{kuar}, \cite{sauder}.
The possibility of using microwave imaging for continuous monitoring of the patient is being considered. The use of microwave radiothermometry to diagnose and monitor progression of cerebral stroke disease and operative counteraction to its uncontrolled growth and possible decision-making support in clinical treatment is discussed by \cite{scapaticci}. Clinical studies show that it is possible to monitor thermal dynamics of biological tissues for several hours. Simulated and experimental results show that a radiometric sensor with a frequency of 1.1-1.6 GHz with a diameter of 2.5 cm is a suitable tool for non-invasive monitoring of brain temperature \cite{rodrigues}. This shows a wide scope of this diagnostic method. In addition to diagnostic purposes, microwave radiothermometry is used to monitor the conditions of laboratory tests. For example, thermal denaturation of albumin by microwave radiometry was monitored in \cite{ivanov}.

Breast cancer modeling for contact thermography and inflammation modeling for the thermo-optical indicator are presented by \cite{biernat}.

A new technique based on the use of electrical impedance to localize preclinical carcinoma emulators in mammary gland agar phantoms \cite {gutierrez} is described. The main idea of the proposed positioning algorithm, called the circular anomaly tracking algorithm, is to find the chest agar region. This region is defined by straight lines connecting pairs of electrodes that have the minimum difference value of a certain normalized impedance along the measurement sweep. This difference was obtained relative to phantom of breast agar without a carcinoma emulator. The proposed technique was evaluated using seven experimental models of agar, six of which had emulators of carcinoma fractions with different locations and electrical conductivity. The authors achieved high rates in the detection of carcinoma.

A separate and independent task is the construction of 3D models of internal and general structure of biological tissues. It must be pointed out, however, that such 3D models are also important for various medical tasks, including preoperative planning and evaluation of oncoplastic surgery \cite{vavourakis}, cosmetic surgery \cite{lopes}, including the establishment of safety criteria for MRI research \cite{park}.

The key role for the radiothermometric method is played by the possibility of increasing its accuracy by improving the design features of the measuring system. The use of receiving antennas with a diameter of 2.5 and 7 cm in the 1.1-1.6 GHz region allows us to track clinically significant changes in deep temperatures for various therapeutic applications \cite{stauffer2}.

The method of microwave radiothermometry is used not only for the diagnosis of paired organs. It is also used in neurology. With any pathology of central and peripheral neural systems, a universal reaction occurs leading to changes in temperature of metabolic, vascular and regulatory genesis. Sustained temperature changes often precede clinical manifestations of pathological process and, therefore, can be a factor in early diagnostics and control of its dynamics.

The main objective of the study is to create a methodology for the initial temperature distribution. This methodology will allow the creation of individual computer models of patients in the future.

\section{Formulation of the problem } \label{sec:datamanagementoverview}

\subsection{Modeling of thermal processes in biological tissues}
The mathematical model is based on three-dimensional non-stationary partial differential equations determining the dynamics of heat. The temperature distributions $ T (\vec {r}, t) $   depend on thermal conductivity coefficient $ \delta (\vec {r}) $, heat sources, metabolic processes $ Q_{met} $, blood flow, cancer formations $ Q_ {car }$ and radiation cooling $Q_ {rad}$. Temperature dynamics is determined by the biothermal equation \cite{pennes}

\begin{eqnarray}
\rho(\vec{r})c_p(\vec{r})\frac{\partial T(\vec{r},t)}{\partial t}=\nabla(\delta(\vec{r})\nabla T(\vec{r},t))+Q_{bl}(\vec{r},t)+ \nonumber
\\
+Q_{met}(\vec{r},t)+Q_{rad}(\vec{r},t),
\end{eqnarray}
where  $\rho$ is substance density, $c_p$ is specific heat of substance, $\nabla=\{\frac{\partial}{\partial x},\frac{\partial}{\partial y},\frac{\partial}{\partial z}\}$.

A boundary condition of the third kind is used in our analysis. In this case, convective heat transfer between the surface of the body and the environment with a constant heat flow is determined as:

\begin{eqnarray}
q=-\delta\frac{\partial T}{\partial \vec{n}}=\alpha(T-T_{air}),
\end{eqnarray}
where $q$ is specific heat flux, $\vec{n}$ is the normal vector, $\alpha$ is heat transfer coefficient,  $T_{air}$ is ambient temperature.

The intensity of heating due to blood flow is controlled by the difference in tissue temperature $ T $, blood $ T_{bl} $, specific heat capacity of blood $ c_{p, b} $

\begin{eqnarray}
Q_{bl}=-\rho\rho_{bl}c_{p,b}\omega_{bl}(T-T_{bl}),
\end{eqnarray}
where $\omega_{bl}$ is blood flow intensity in heating region, the values of which can lie in very wide limits $\omega_{bl}=4\cdot 10^7 - 2\cdot 10^{-5}$ m${^3}$/kg$\cdot$ s.

Among other heat sources we can mention: heat generation in tissues as a result of vital processes, blood flow, which is the most important source of thermal energy and radiation re-emission.
Also, sources due to cancer cells should be taken into account.
The specific energy density is determined by intensity of biochemical
processes in tissues. Its typical values are 4000--5000 W/m$^ 3$.
The temperature in biological tissues changes significantly in the presence of tumors.

Earlier, numerical experiments were carried out to reveal dependence of temperature dynamics on various parameters of biological tissue \cite{polyakov}. It was shown that the dynamics of temperature depends strongly on parameters of biological components.

An important problem in the method of microwave thermometry is determination of the brightness temperature, which is different from thermodynamic temperature. This temperature is measured using microwave antennas.
The antenna with a frequency of several GHz allows us to measure thermal radiation from biological tissues in a certain frequency range.

Brightness temperature infers from the following equation
\begin{eqnarray}
 T_B(\vec{r}) = \int\limits_{\Delta{f}} \Biggl\{ \left| S_{11}(f) \right|^2 T_{REC} + \left[ 1 - \left| S_{11}(f) \right|^2 \right] \times
 \nonumber
 \\
 \times  \left( \int_{V_{0}} T(\vec{r}) \,\frac{P_d(\vec{r},f)}{\int_{V_0}P_d(\vec{r},f)\,dV}\,dV + T_{EMI} \right) \Biggr\}\, df \,,
 \label{eq-TBintegral}
\end{eqnarray}
where $\displaystyle P_d = \frac{1}{2}\,\sigma(\vec{r},f)\cdot \left| \vec{E}(\vec{r},f) \right|^2$ is the electromagnetic field power density, $\vec{E}$ is electric field vector. Values $T_{EMI}$ and $T_{REC}$ characterize electromagnetic interference when measured with a radiometer \cite{vesnin}. Coefficient $S_{11}$ determines the interaction between the antenna and the biological tissue. Integration is carried out over the entire volume of biotissue ($V_0$).

To construct a stationary electric field distribution, it is convenient to solve the time-dependent Maxwell equations and as the result to obtain the stationary state:
\begin{eqnarray}
    \frac{\partial \vec{B}}{\partial t} + rot(\vec{E}) = 0 \,, \quad \frac{\partial \vec{D}}{\partial t} - rot(\vec{H}) = 0 \,,\quad \vec{B}=\mu\vec{H}\,,\nonumber
    \\
    \quad \vec{D}=\varepsilon\vec{E} \,,\;\;\;\;
\end{eqnarray}
where  $\vec{B}$ is magnetic induction, $\vec{E}$ is electric field strength, $\vec{D}$ is electric induction, $\vec{H}$ is magnetic field strength, $\varepsilon(\vec{r})$ is the dielectric constant, $\mu(\vec{r})$ is magnetic permeability.

\subsection{ An object to be simulated and its geometry}
The female breast is an organ with a complex organization, the structure of which
should create the most optimal conditions for fulfilling its main
physiological functions (production of milk and feeding of a child).

\begin{figure} [h!]
 \caption{ Schematic structure of the breast \cite{peterson} }
\centerline{\includegraphics[scale=0.22]{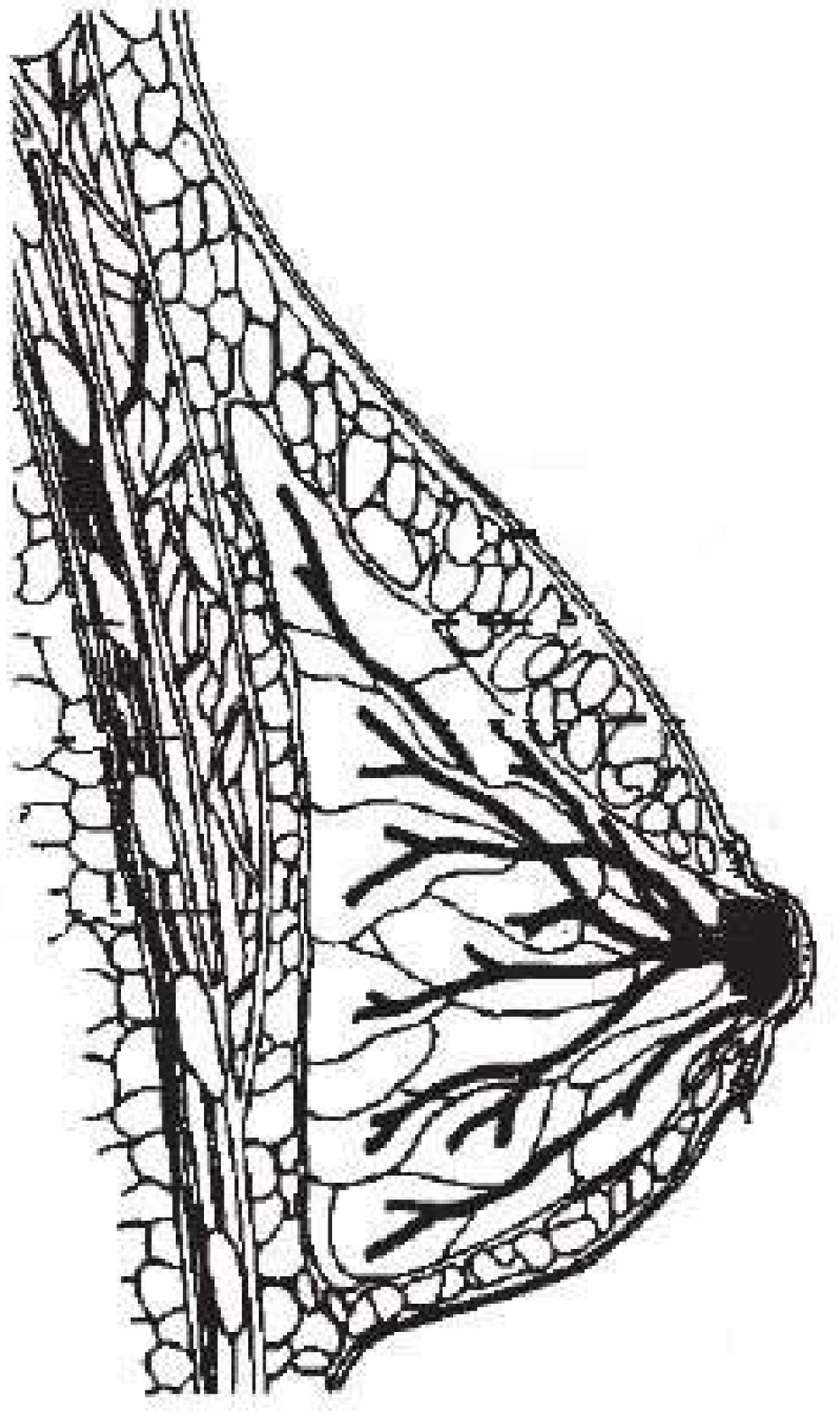}}
\label{Structure}
\end{figure}

Breast consists of a layer of skin under which the milk gland is located
(glandular tissue) -- it is in this organ milk is produced. The mammary gland is attached with the help of connective tissues to muscles of the chest (see Figure \ref{Structure}). The mammary gland consists of 15--20 glandular lobules, adipose and connective tissue. The sizes of all components vary from one person to another.

\begin{figure} [h!]
 \caption{ The projection of the model’s geometry on a 2D plane }
\centerline{\includegraphics[scale=0.22]{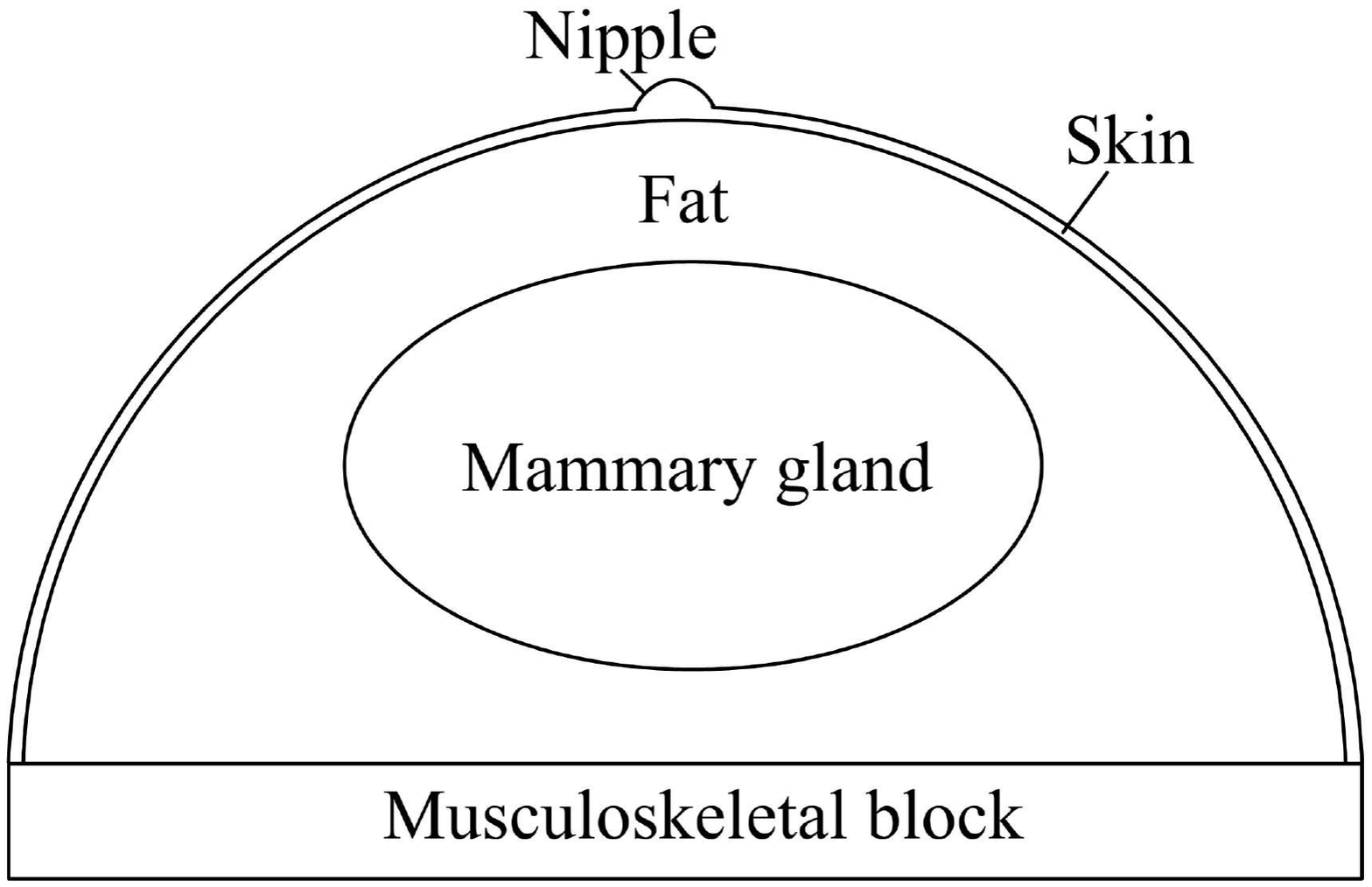}}
\label{Geometry}
\end{figure}

The mammary gland is located on the pectoralis major muscle. The mammary gland is covered with very thin skin, which easily moves over
base. Under the skin is a fat layer, the thickness of which can vary from one woman to another. Under the layer of fat is the body of the mammary gland, covered with connective tissues
a capsule by which it is suspended from collarbone. Most of the breast is filled with fat. Amount of fat in female breasts varies significantly. For some women
their breast consists almost exclusively of fat, while the others have glandular tissue
in the breast covers more space than fat. Figure \ref{Geometry} shows simplified model geometry.

\subsection{ A method to measure the internal temperature of the mammary glands }
Currently used complex RTM-01-RES allows us to evaluate the functional state of tissues by measuring their internal temperature (RTM) at a depth of 5 cm and skin temperature (IR). Examination of a patient is carried out in a horizontal position, naked to waist, hands under the head. The examined area 15 minutes before the measurement is released from clothes for acclimatization to room temperature of the entire measurement area. Due to the presence of radio noise in atmosphere and to exclude influence of position antenna in space on the measurement results, latter
during examinations, it is recommended to orient the patient in one direction. Thus,
when measuring thermal emission of symmetrical points, the patient changes position, sitting on
swivel chair. The receiving antenna without air gaps is pressed against the skin surface above the  temperature measurement area. After stabilization of the parameters, that software monitors and confirms, measured temperature entered into the database.

\begin{figure}[h!]
 \caption{ The scheme for measuring the temperature of the mammary glands \cite{zenovich}}
\centerline{\includegraphics[scale=0.25]{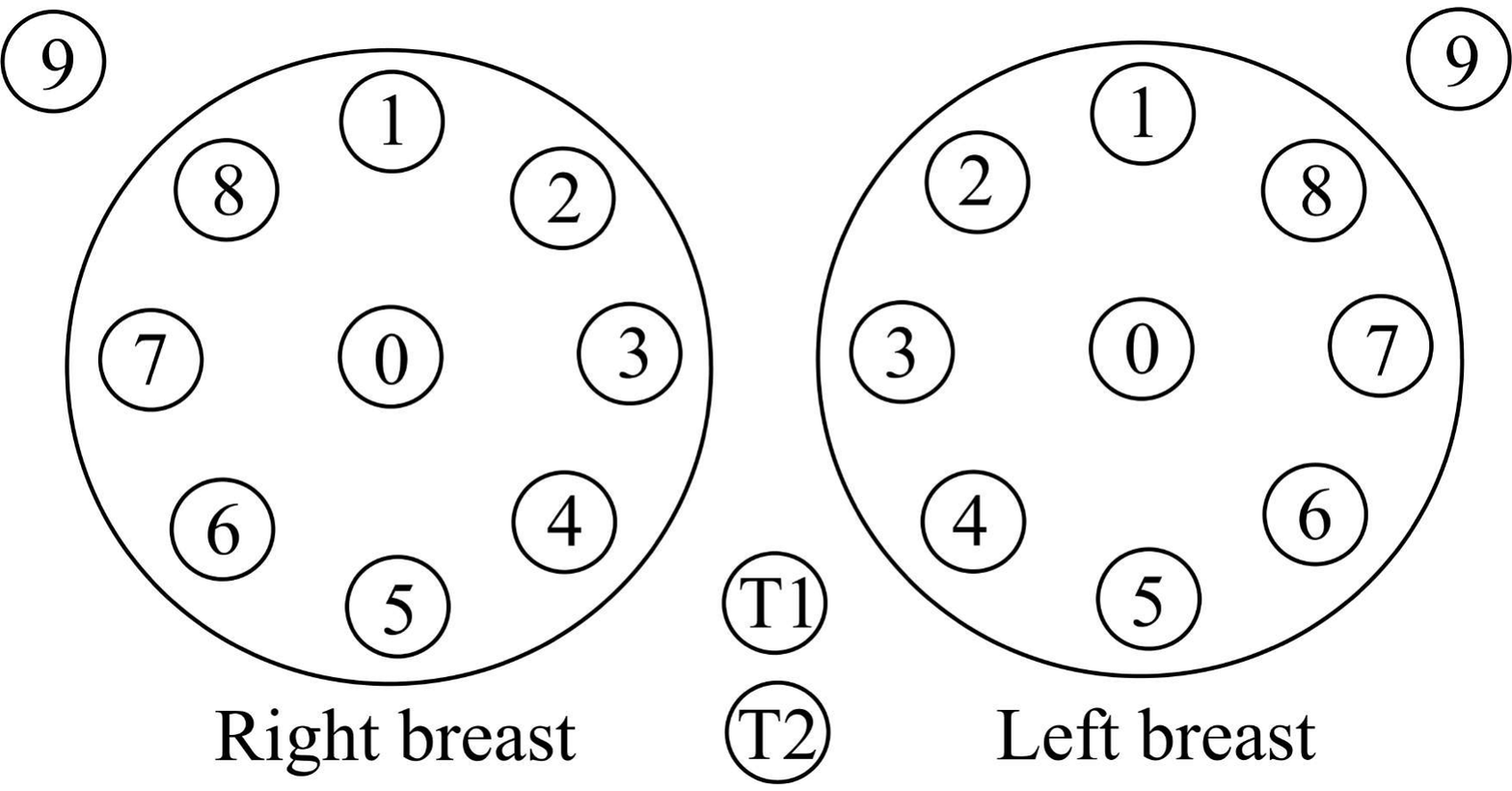}}
\label{Scheme}
\end{figure}


The examination begins with measuring the temperatures at the reference points T1 and T2. The first point is located in the center of the breast immediately below and between the mammary glands; the second is located directly under the xiphoid process. Further measurements are carried out at 10 points on each gland and in the axillary region (see Figure \ref{Scheme}). In accordance with the methodology, measurements should be carried out at an ambient temperature of 20 to 25 degrees.

\subsection{ Description of the problem }
The construction of simulation models for the mammary glands will allow us to evaluate temperature anomalies and detect structural changes in internal tissues. There is a need to solve the inverse problem of thermal conductivity to develop such models. This would enable us  to restore the structure of the mammary gland according to the known distribution of deep and skin temperatures.

\begin{figure}[h!]
 \caption{The procedure for finding the coefficient of thermal conductivity}
\centerline{\includegraphics[scale=0.15]{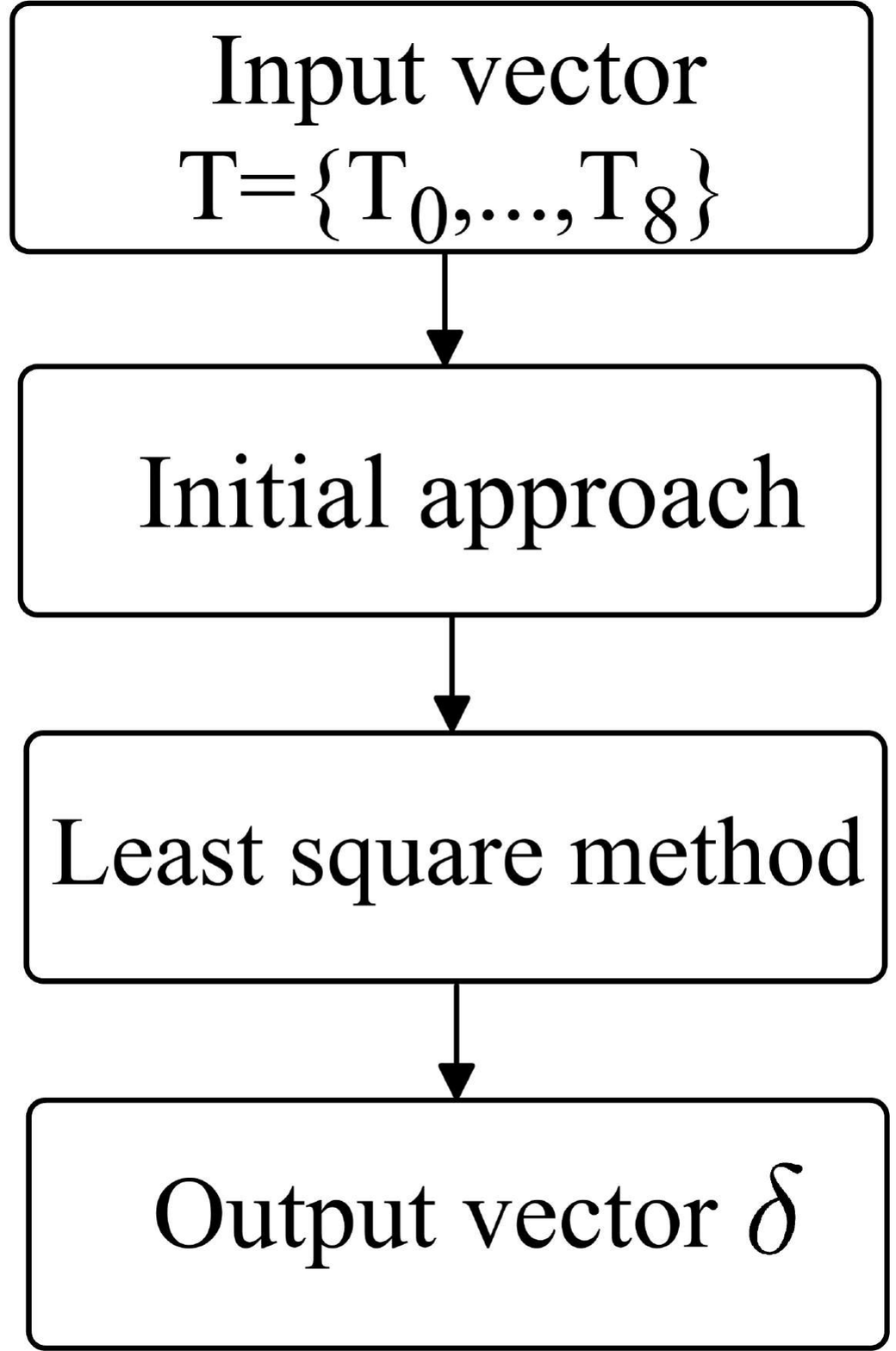}}
\label{Algorithm}
\end{figure}

As input parameters, we will use the temperature data vector $\vec{T}=\{T_0, ..., T_{8} \}$ for one mammary gland the right gland was chosen for this obtained using a computational experiment in the microwave in the infrared range. At an output, it is necessary to obtain the thermal conductivity coefficients of various biological tissues with other parameters being fixed. Figure \ref{Algorithm} shows the procedure for finding this coefficient.

\section{ Numerical solution of the inverse task using optimization methods }
\subsection{ Description of method }
The process is assume to be transient. At the boundary, a constant ambient temperature is set. The value of this temperature, $ T_ {air} $, is taken from experimental data. The values of predicted $ \vec{T} = \{T_0, ..., T_ {8} \}$, and measured temperatures are compared using the minimizing of the quadratic residual functional

\begin{eqnarray}
A=\sum_{i}(T_i-T^{exp}_i)^2 \rightarrow {\delta}_{min}.
\end{eqnarray}

The objective function is the sum of squares of the differences between the measured $ T ^ {exp} _i $ and the calculated temperature values $ T_i $. The control parameter is thermal conductivity coefficient $\delta $.
It is required to find a vector $ \vec {\delta} $ that minimizes the discrepancy $ N $. Due to the incorrectness of the problem, we minimize the Tikhonov functional. Moreover, the functional is defined positively, therefore, it has a single minimum.

\begin{figure}[h!]
 \caption{ Initial temperature distribution }
\centerline{\includegraphics[scale=0.23]{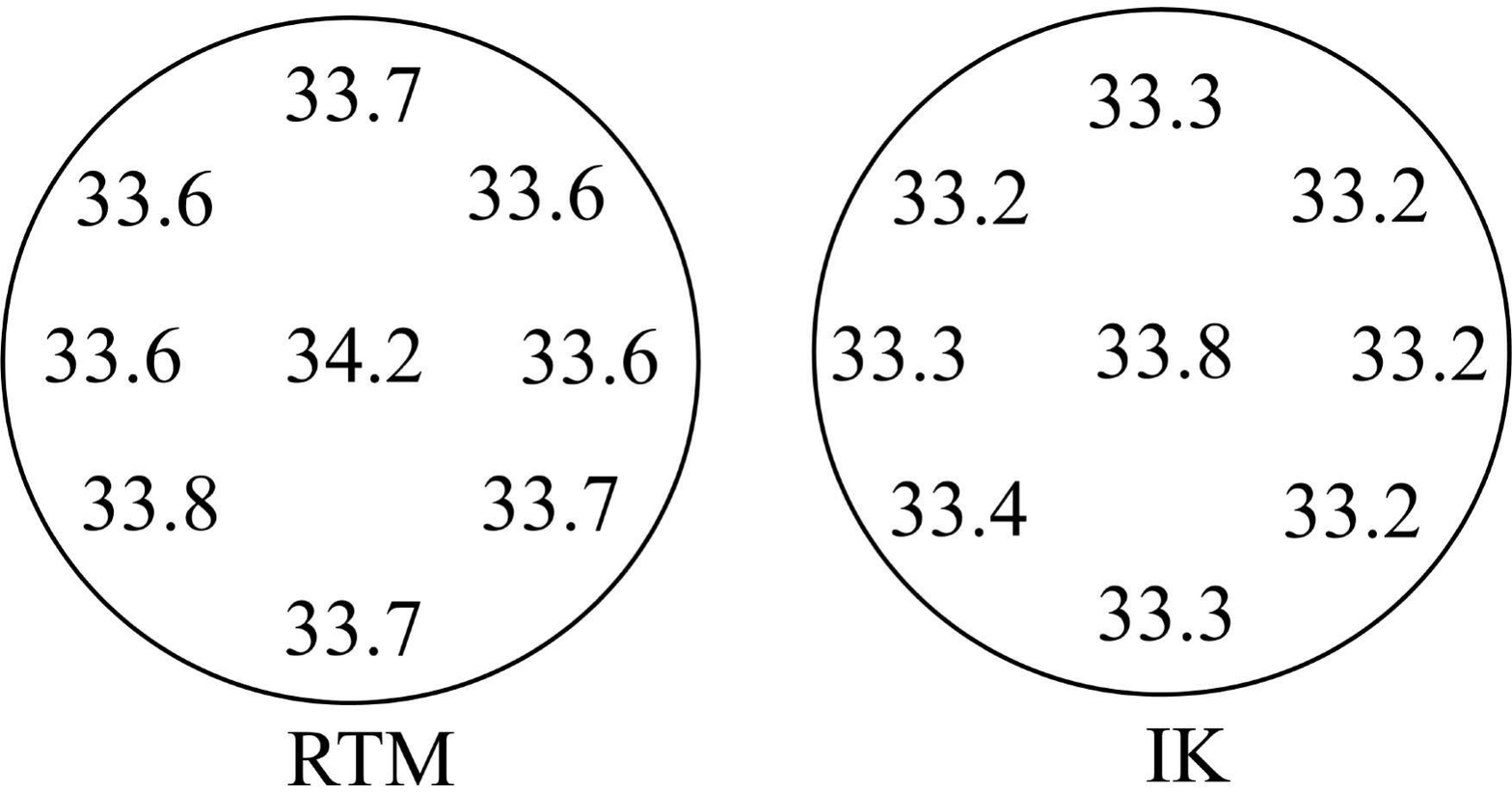}}
\label{Initial}
\end{figure}

\subsection{ Computational Experiment }
The objective function is set as the vector of temperatures measured in the microwave range $\vec{T}_{RTM}=\{34.2, 33.7, 33.6, 33.6, 33.7, 33.7, 33.8, 33.6, 33.6\}$ and in the infrared range $\vec{T}_{IK}=\{$33.8, 33.3, 33.2, 33.2, 33.2, 33.3, 33.4, 33.3, 33.2$\}$ (see Figure \ref{Initial}).
The temperatures are set in $^o$C in all cases.

\begin{table}[h!]
\center{{\bf Table 1.} Results of a numerical experiment with control parameter $\delta$}
\begin{center}
\normalsize
\begin{tabular}{|l|c|c|}
\hline
Biotissue & \begin{tabular}[c]{@{}c@{}}Calculated value\\ W/(m $^\circ$C)\end{tabular} & \begin{tabular}[c]{@{}c@{}}Exact value\\ W/(m $^\circ$C)\end{tabular} \\ \hline
$\delta_{skin}$      & 0.42                                                          & 0.45                                                     \\ \hline
$\delta_{fat}$       & 0.24                                                          & 0.2                                                      \\ \hline
$\delta_{mam.gl.}$  & 0.41                                                          & 0.4                                                      \\ \hline
$\delta_{nipple}$    & 0.44                                                          & 0.4                                                      \\ \hline
\end{tabular}
\end{center}
\end{table}

After performing $ \sim 10^8$ iterations, the following thermal conductivity vector was obtained $ \vec {\delta^*} = \{0.42, 0.24, 0.41, 0.44 \}$. The calculated and measured values of thermal conductivity coefficients differ by about $ \sim 10 ^{- 2}$ (Table 1).

\begin{figure} [h!]
 \caption{ Verification of the results of a computational experiment (dashed line) with an exact solution (solid line)}
\centerline{\includegraphics[scale=0.27]{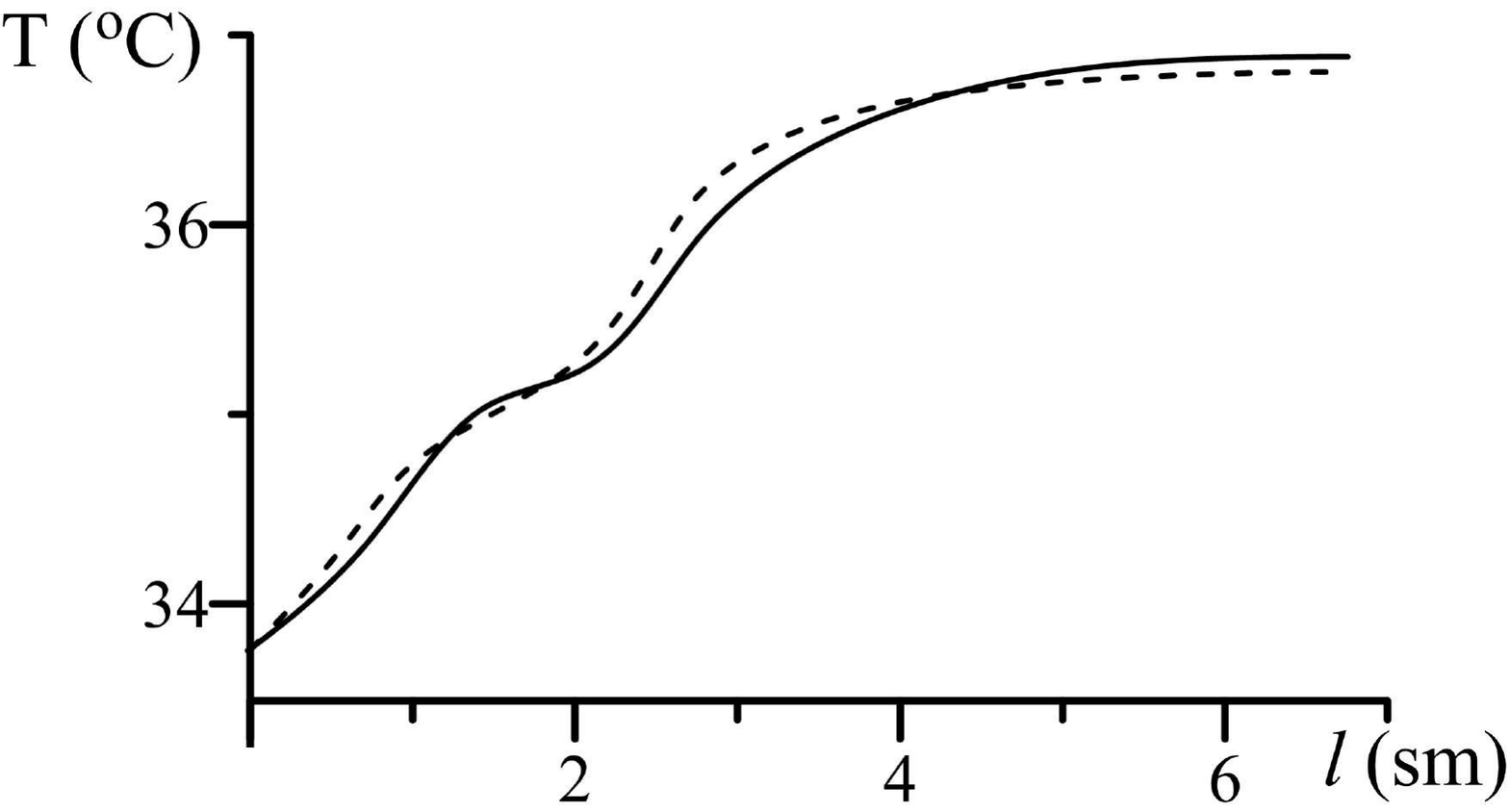}}
\label{Temperature}
\end{figure}

Remembering that the structure is very inhomogeneous, this result completely satisfies the requirements of the task. At the next stage we conducted a direct computational experiment to verify the results and use the resulting vector $\vec{\delta^*} $ as input parameters. The resulting temperature distribution was constructed along the axis of axial symmetry (see Figure \ref{Temperature}).

We carried out a similar procedure, but fixed the thermal conductivity coefficients of biological tissues, and made the specific heat $ c_p $ a controlling parameter.

\begin{table}[h!]
\center{{\bf Table 2.} Results of a numerical experiment with a control parameter $c_p$}
\begin{center}
\normalsize
\begin{tabular}{|l|c|c|}
\hline
Biotissue & \begin{tabular}[c]{@{}c@{}}Calculated value\\ J/(kg $^\circ$C)\end{tabular} & \begin{tabular}[c]{@{}c@{}}Exact value\\ J/(kg $^\circ$C)\end{tabular} \\ \hline
${c_p}_{skin}$      & 3100.7                                                          & 3100                                                  \\ \hline
${c_p}_{fat}$       & 2599.4                                                         & 2600                                                      \\ \hline
${c_p}_{mam.gl.}$  & 3199.2                                                          & 3200                                                      \\ \hline
${c_p}_{nipple}$    & 3000.8                                                         & 3000                                                      \\ \hline
\end{tabular}
\end{center}
\end{table}

Table 2 shows that error in determining the specific heat is $ \sim $ 0.1, which is an acceptable result of numerical experiment.

The experimental results indicate that the proposed method gives fairly good results and provides the necessary accuracy of solution as a first approximation.

\subsection{Parallel technologies }
The process of solving the problem is resource-intensive. It makes sense to use parallel computing technologies to reduce the time spent on calculations. The easiest way to parallelize the code is to use OpenMP standard for multithreaded software for systems with shared memory, since in this case specialized computer servers are not required, and the application is rather universal in use.

\begin{figure}[h!]
 \caption{ Dependence of the acceleration of the parallel version relative to the serial version. The graph shows the dependence of the acceleration on the number of iterations.}
\centerline{\includegraphics[scale=0.25]{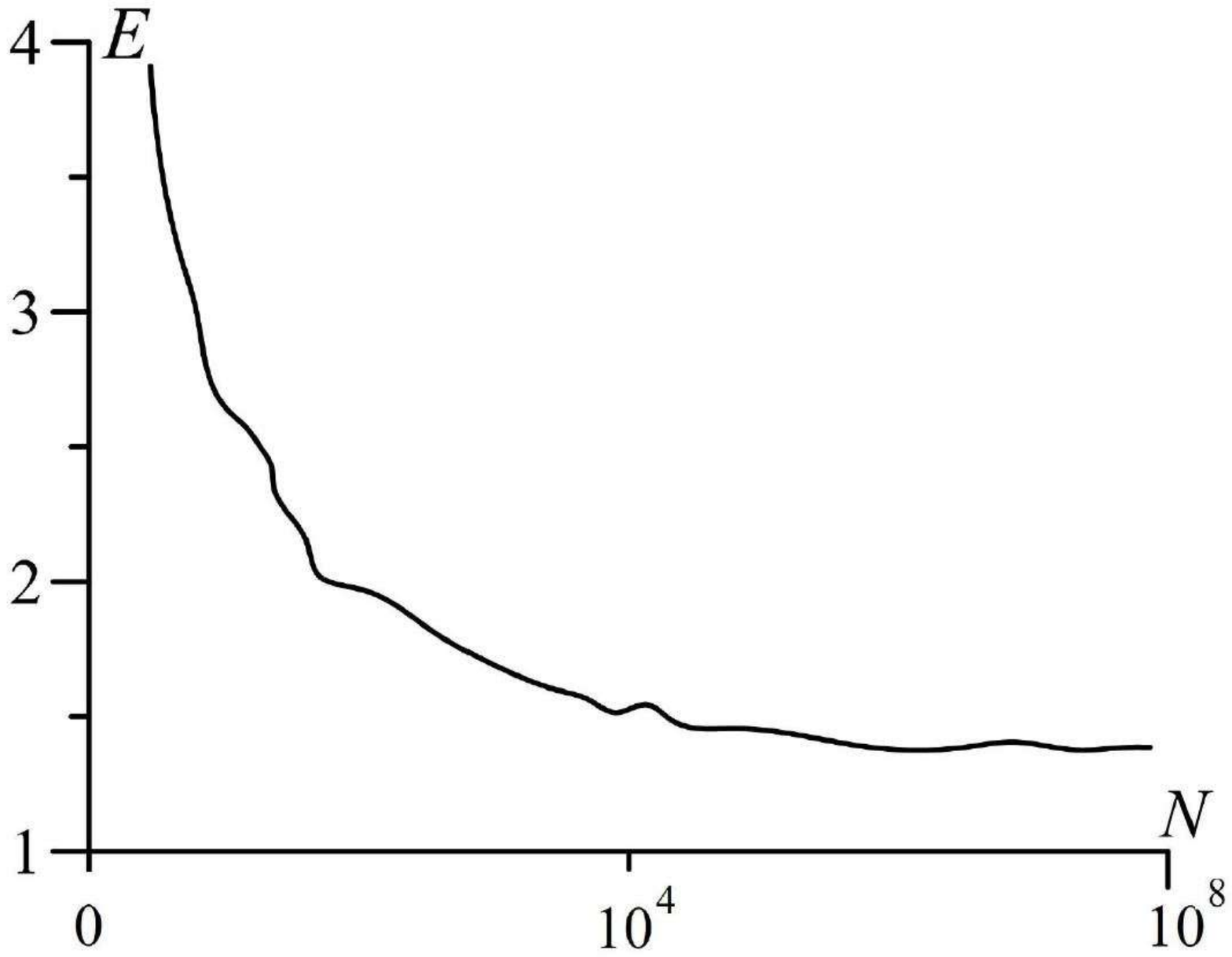}}
\label{Parallel}
\end{figure}

Tasks performed by threads in parallel, as well as the data required to perform these tasks, are described using special preprocessor directives of the corresponding language pragmas. The number of created threads can be controlled both by the program itself by calling library procedures, and from the outside using environment variables. As follows from the comparison of the results, the OpenMP version gives an average acceleration of 1.8 times compared to the serial version (see Figure \ref{Parallel}).

\section{ Discussion of the results }

Matching the results of computational experiments with
laboratory experiments allows us to get more accurate physico-chemical characteristics of various tissues and components of the mammary gland. It is expected that this will increase efficiency of timely diagnose of tumors.

A serious problem for diagnostics and modeling of this phenomenon lies in a wide rage of basic parameters characterizing tissues in various organisms (thermal conductivity, transfer and scattering coefficients, electrical conductivity, dielectric constant, heat capacity, blood viscosity, blood flow parameters, heat transfer of the capillary system). The situation is significantly complicated due to strong heterogeneity of the mammary glands with characteristic spatial scales of less than 1 cm. The spread in the characteristics of the tumor tissue can malignant tumor tissue reach order. For example, strong dependence of the heat release on its doubling time is well known. It is possible to use cluster computing to solve  large
distributed memory models. To speed up work,
cluster implementation in software can
apply multi-core computing with shared memory in each node
combined with an interface-based distributed memory model
message passing MPI (Message Passing Interface). This approach, known as hybrid concurrency, speeds up the work considerably due to efficient use of computing resources.

\section{ Conclusion }
A method for modeling the dynamics of thermal processes in biological tissues of the mammary gland has been developed. This method provides the required accuracy of solutions, stability and the high convergence rate required by personalized medicine.
The inverse problem of reconstructing thermal conductivity coefficient in the equation of heat dynamics in biological tissues from the final temperature distribution  has been solved. This solution is used in the mathematical model for determining the temperature distribution inside the mammary gland.
A numerical example of solving the inverse problem has been studied in details.

The lack of detailed qualitative and quantitative description of the behavior of temperature fields in various human organs, both in the presence of pathological processes and in their absence, significantly complicates the development of
effective methods of medical diagnosis.
Our approach is the first step towards solving this problem. The model of mammary gland was presented as quasi-one-sided, and minor details of the mammary gland were not taken into account.

\section*{Acknowledgement}
The reported study was funded by RFBR according to the research project No. 19-37-90142.



\begin{thebibliography}{10}

\bibitem{antonova}
Antonova,  M., Totev, T., Antonova, S., Zahariev, T. and Stoytchev, S. (2016) `A new device for static and dynamic investigation of the in vitro viscoelastic characteristics of biological tissues', {\it Comptes rendus de l'Academie Bulgare des Sciences}, Vol. 69, No. 4, pp. 497--504.

\bibitem{barrett}
Barrett, A.H. and  Myers, P.C.  (1975) `Subcutaneous temperature: a method of noninvasive sensing', {\it Science.}, Vol. 190, pp. 669--671.

\bibitem{biernat}
Biernat, J., Biernat, M., Lukasik, W., Palko, T., Jung, A. and Trzyna, M. (2019) `Physical breast model as a simulator of pathological changes', {\it World Congress on Medical Physics and Biomedical Engineering}, Vol. 68, No. 1, pp. 795-798.

\bibitem{bressan}
Bressan, R.S., Bugatti, P.H. and Saito, P.T.M.  (2019) `Breast cancer diagnosis through active learning in content-based image retrieval', {\it Neurocomputing}, Vol. 357, pp. 1--10.

\bibitem{gomes}
Figueiredo, A. A. A., do Nascimento, J. G., Malheiros, F. C., da Silva Ignacio, L. H., Fernandes, H. C., Guimaraes, G. (2019). Breast tumor localization using skin surface temperatures from a 2D anatomic model without knowledge of the thermophysical properties, {\it Computer Methods and Programs in Biomedicine}, Vol. 172, pp. 65--77.

\bibitem{khokhlova}
Filonenko, E.A. and Khokhlova, V. (2000) `Modelling of thermal processes in biological tissues under the action of focused ultrasound', {\it Russian Ultrasonics}, No. 30, pp. 13--19.

\bibitem{gautherie1}
Gautherie, M. and  Gros, C.M. (1980) `Breast thermography and cancer risk prediction', {\it Cancer}, Vol. 45, pp. 51--56.

\bibitem{gautherie2}
Gautherie, M. (1982) `Temperature and blood flow patterns in breast cancer during natural evolution and following radiotherapy', {\it Prog. Clin. Biol. Res.}, No. 107, pp. 21--64.

\bibitem{gupta}
Gupta, N. K., Srivastava, S.K. and Tiwari, H.V.  (2004) `Estimation of emissivity characteristics of biological tissues at microwave frequency', {\it IE (I) Journal-ID}, Vol. 84, pp. 1-3.

\bibitem{gutierrez}
Gutierrez-Lopez, M., Prado-Olivarez, J., Diaz-Carmona, J., Herrera-Ramírez, C., Gutierrez-Gnecchi, J. and Medina-Sanchez, C. (2019) `Electrical impedance-based methodology for locating carcinoma emulators on breast models', {\it Journal of Sensors}, Vol. 2019, pp. 1--16.

\bibitem{ivanov}
Ivanov, Y.,  Kozlov, A., Galiullin, R., Tatur, V., Ziborov, V., Ivanova, N., Pleshakova, T., Vesnin, S. and Goryanin,  I. (2018) `Use of microwave radiometry to monitor thermal denaturation of albumin', {\it Frontiers in Physiology}, Vol. 9, pp. 1--11.

\bibitem{kosin}
Kosin, I. U.,  Leonidov, I.V.,  Kravzov, V. A. and Bobne, A. R. (2017) `Device for measurement of biological tissue characteristics', {\it Telecommunications and Radio Engineering.}, No. 76, pp. 1173--1179.

\bibitem{kuar}
Kaur, R.P.,  Kumar, V., Shafi, G., Vashistha, R., Kulharia, M. and Munshi, A. (2019) `A study of mechanistic mapping of novel SNPs to male breast cancer', {\it Medical Oncology}, Vol. 36, No. 8. 70.

\bibitem{liu}
Li, Y., Sun, C.,  Kuang, J.J., Ji, C.C., Li, J., and Wu, J.T. (2018) `Experimental study on heat transfer characteristics of biological tissue during moxibustion therapy', {\it Journal of Engineering Thermophysics}, No. 39, pp. 150--154.

\bibitem{lopes}
Lopes, D., Clain, S., Pereira, R.M.S., Machado , G.J., Smirnov, G. and Vasilevskiy, I. (2017) `Numerical simulation of breast reduction with a new knitting condition', {\it  Int. J. for Numerical Methods in Biomedical Engineering}, Vol. 33, No. 2,  e02796.

\bibitem{losev}
Losev, A.G., Levshinskiy, V.V. (2017)  `Data mining of microwave radiometry data in the diagnosis of breast cancer', {\it Mathematical Physics and Computer Simulation}, Vol. 20, No. 5. pp. 49--62.

\bibitem{park}
Park, S.O., Hwang, J.H., Kang, T.W. and Kwon, J.H. (2017) `Effect of electromagnetic interference on human body communication', {\it  IEEE Transactions on Electromagnetic Compatibility}, Vol. 59, No. 1, pp. 48--57.

\bibitem{pennes}
Pennes, H.H. (1948)  `Analysis of tissue and arterial blood temperature in the resting human foream', {\it  J. of Appl. Physiology}, Vol. 40, pp. 24--30.

\bibitem{peterson}
Peterson, B.E., Chissov, A.I. and Paches, A.I. (1987) `Atlas of cancer operations', {\it  Medicine}, pp. 532--534.

\bibitem{polyakov}
Polyakov,  M.V., Khoperskov,  A.V. and Zamechnic,  T.V. (2017) `Numerical modeling of the internal temperature in the mammary gland', {\it  LNCS}, Vol. 10594, pp. 128--135.

\bibitem{rodrigues}
Rodrigues, D.B., Stauffer,  P.R., Pereira,  P.J.S. and Maccarini, P.F. (2018) `Microwave radiometry for noninvasive monitoring of brain temperature', {\it Crocco L., Karanasiou I., James M., Conceicao R. (eds) Emerging Electromagnetic Technologies for Brain Diseases Diagnostics, Monitoring and Therapy. Springer, Cham}, pp. 87--127.

\bibitem{rotter}
Rotter, J., Wilson, L., Greiner, M.A., Pollak, C.E. and Dinan, M. (2019) `Shared-patient physician networks and their impact on the uptake of genomic testing in breast cancer',  {\it Breast Cancer Research And Treatment}, Vol. 176, No. 2, pp. 445--451.

\bibitem{saady}
Saady, A., Bottner, V., Meng, M., Varon, E., Shav-Tal, Y., Ducho, C. and Ficher, B. (2019) `An oligonucleotide probe incorporating the chromophore of green fluorescent protein is useful for the detection of HER-2 mRNA breast cancer marker', {\it European Journal Of Medicinal Chemistry}, Vol. 173, pp. 99--106.

\bibitem{sarvazyan}
Sarvazyan, A. (1975) `Low-frequency acoustic characteristics of biological tissues', {\it Mechanics of Composite Materials }, No. 11, pp. 594--597.

\bibitem{sauder}
Sauder, C.A.M. (2019) `ASO author reflections: surgical treatment for male breast cancer in the modern era', {\it Annals Of Surgical Oncology}, Vol. 26, No. 7, pp. 2154--2155.

\bibitem{scapaticci}
Scapaticci, R.,  Bucci, O.,  Catapano, I. and  Crocco, L. (2014) `Differential microwave imaging for brain stroke followup', {\it International Journal of Antennas and Propagation}, Vol. 2014, No. 6, pp. 1--11.

\bibitem{sedankin}
Sedankin, M.K., Leushin, V.Y., Gudkov, A.G., Vesnin, S.G., Sidorov, I.A., Agasieva, S.V., Ovchinnikov,
L.M. and Vetrova, N.A. (2018) `Antenna applicators for medical microwave radiometers', {\it Biomedical Engineering}, Vol. 52, No. 4, pp. 235--238.

\bibitem{sharma}
Sharma, R. (2019) `Breast cancer incidence, mortality and mortality-to-incidence ratio (MIR) are associated with human development, 1990–2016: evidence from Global Burden of Disease Study 2016', {\it Breast Cancer}, Vol. 26, No. 4, pp. 428--445.

\bibitem{sheeba}
Sheeba, I.R. and Jayanthy, T. (2019) `Design and analysis of a flexible softwear antenna for tumor detection in skin and breast model', {\it Wireless Personal Communications}, Vol. 107, No. 2, pp. 887--905.

\bibitem{shirmohammadli}
Shirmohammadli, V. and Manavizadeh, N. (2018) `Numerical modeling of cell trajectory inside a dielectrophoresis microdevice designed for breast cancer cell screening', {\it IEEE Sensors Journal}, Vol. 18, No. 20, pp. 8215--8222.


\bibitem{stauffer}
Stauffer, P., Rodrigues, D., Salahi, S., Topsakal, E., Oliveira, T., Prakash, A., D'Isidoro, F., Reudink, D., Snow, B. and Maccarini, P. (2013) `Stable microwave radiometry system for long term monitoring of deep tissue temperature', {\it Proceedings of SPIE}, Vol. 8584, pp. 1--18.

\bibitem{stauffer2}
Stauffer, P.R. and Rodrigues, D.R. (2014) `Utility of microwave radiometry for
diagnostic and therapeutic applications of non-invasive temperature monitoring',  {\it  IEEE
BenMAS (Benjamin Franklin Symposium on Microwave and Antenna Sub-systems)}, Vol. 2014, pp. 1--3.

\bibitem{sundell}
Sundell, V., Jousi, M., Hukkinen, K., Blanco,  R., Makela, T. and Kaasalainen, T.  (2019) `A phantom study comparing technical image quality of five breast tomosynthesis systems', {\it European Journal of Medical Physics}, Vol. 63, pp. 122--130.

\bibitem{vavourakis}
Vavourakis, V., Eiben, B., Hipwell, J.H., Willams, N.R., Keshtgar, M. and Hawkes, D.S. (2016) `Multiscale mechano-biological finite element modelling of oncoplastic breast surgery—numerical study towards surgical planning and cosmetic outcome prediction', {\it PloS one}, Vol. 11, No.7, e0159766.

\bibitem{vesnin}
Vesnin, S., Turnbull, A.,  Dixon, M.J. and Goryanin, I. (2017) `Modern microwave thermometry for breast cancer', {\it Molecular and cellular biomechanics: MCB}, Vol. 7, No. 2, e.1000136.

\bibitem{yahyazadeh}
Yahyazadeh , S. and Mehraeen, R. (2019) `A comparison of the diagnostic value of magnetic resonance mammography versus ultrasound mammography in moderate- and high-risk breast cancer patients', {\it J. Evolution Med. Dent. Sci.}, Vol. 7, No. 53, pp. 5629--5633.

\bibitem{zaheditochai}
Zaheditochai, M.,  Jaafar, R. and Zahedi, E. (2007) `Non-invasive techniques for assessing the endothelial dysfunction: ultrasound versus photoplethysmography', {\it 13th Biomedical Engineering}, Vol. 23, pp. 65--68.

\bibitem{zenovich}
Zenovich, A.V., Baturin, N.A., Medvedev, D.A. and Petrenko, A.Y. (2018) `Algorithms for
the formation of two-dimensional characteristic and informative signs of diagnosis
of diseases of the mammary glands by the methods of combined radiothermometry', {\it Math. Phys. Comput. Simul.}, Vol. 21, No. 4. pp. 44--56.

 \end{thebibliography}
\end{document}